\documentclass{article}
\usepackage{cite}
\usepackage{amsfonts,amssymb,amsmath}
\allowdisplaybreaks[1]
\let\kappa=\varkappa
\newcommand{\°}{{\!}}
\newcommand{\Sr}{\;\! ; \;\!}
\newcommand{\Sp}{\;\! , \;\!}
\newcommand{\abc}[1]{\mbox{#1)}\quad}
\newcommand{\bm}[1]{\boldsymbol{#1}}
\newcommand{\ds}{\displaystyle}
\newcommand{\dd}{{\mathrm{d}}}
\newcommand{\deriv}[2]{\,\mbox{$\displaystyle \dfrac{{\rm d}#1}{{\rm d}#2}$}\,}
\newcommand{\sig}{\delta}

\newcommand{\GN}{\gamma_{\mathrm{N}}}
\newcommand{\Deriv}[2]{\mbox{$\displaystyle \frac{{\rm D}#1}{{\rm D}#2}$}}
\newcommand{\pderiv}[2]{\mbox{$\displaystyle\frac{{\partial}#1}{{\partial}#2}$}}
\newlength{\temp}
\newcommand{\pot}[1]{\cdot 10^{#1}}
\newcommand{\PHI}{\Phi_{\mathrm{eff}}}
\newcommand{\kk}[1]{\,\mathrm{#1}}

\newcommand{\so}{\Bar{\sigma}{}}
\newcommand{\si}{s}
\newcommand{\mut}{\mu^{(\sigma)}}

\begin{document}

  \title{FIELD EQUATIONS AND EQUATIONS OF MOTION IN POST-NEWTONIAN
APPROXIMATION OF THE PROJECTIVE UNIFIED FIELD THEORY\footnote{Dedicated to
Prof. Dr. Nikolay Mitskievich  on the occasion
of his $80^{\rm th}$ birthday.}}

\author{ALEXANDER GORBATSIEVICH\\[1ex]
Department of Physics of Belorussian State University, Nesavisimosti av., \\[1ex]
Minsk, 220030, Belarus\\
Gorbatsievich@bsu.by\\[2ex]
ERNST SCHMUTZER\\[1ex]
Friedrich Schiller University of Jena\\
D-07743 Jena, Germany\\
eschmutzer@t-online.de}
\date{}

\maketitle

\begin{abstract}
  The equations  of motion of $N$ gravitationally bound  bodies are  derived  from
         the field equations of Projective Unified Field Theory. The Newtonian  and  the  post-Newtonian
         approximations of the field  equations  and  of  the  equations  of
         motion of this system of bodies are studied in  detail.  In  analyzing  some
         experimental data we performed some numeric estimates of the ratio
         of the inertial mass to the scalaric mass of matter.
\end{abstract}
\maketitle
\section{Introduction}
Recently considerable interest raised by experiments aiming at the
verification of the relativistic theories of gravitation (in detail
see \cite{Will} and the literature quoted there). The high precision
experiments on checking the validity of the equivalence principle,
perihelion shift of Mercury, deflection of light and other effects
give some answer to the question of preference among further
gravitational theories proposed, where the distinction between them
is mainly based on the post-Newtonian or even on the
post-post-Newtonian level.

Among the alternative theories of gravitation here our special
interest is directed to the 5-dimensional Projective Unified Field
Theory of Schmutzer (short: PUFT). For details of the axiomatics as
well as the cosmological and astrophysical application of PUFT see
\cite{Schmutzer1,Schmutzer2,SchmuGru} and \cite{Gor1}. Within the framework
of this theory recently interesting cosmological models were
investigated \cite{Schmutzer3,SchmuGru}, where in particular an
unexpected solution of the problem of dark matter was suggested.
For some observable cosmological effects see also \cite{Blin}.
Obviously within the framework of PUFT the systematic study and
subsequent empirical testing of various astrophysical effects is
necessary, particularly effects in the solar system. First steps in
this direction were made in the papers of E. Schmutzer quoted above.
However the wishes of astrophysicists, particularly of those being
active researchers in celestial mechanics, ask for a systematic
treatment of this problem by building up the post-Newtonian
approximation of PUFT.

Schmutzer's PUFT has been developed in three stages: The version I
of PUFT (1958) led to a possible violation of the equivalence
principle which nowadays has already been tested experimentally to a
relative precision of $10^{-12}\div 10^{-14}$ \cite{Will}.
Two decades later this situation was one of the reasons for him to
elaborate the version II of PUFT (GR9-conference in Jena 1980),
where a projection formalism with a kind of ``conformal projection
factor'' was used. Emphasizing his   5-dimensional concept of describing real
physics, intended  since the beginning of his research in this field (see \cite{Schmuhab}), he already then introduced the concept of scalarism (scalarity) as a
hypothetical new phenomenon of Nature: PUFT unifying gravitation,
electromagnetism and scalarism.

Detailed information on this historical development of PUFT and present state of PUFT can also
be found in the monographs \cite{Schmutzer2,SchmuGru} and \cite{Schmutzer2,SchmuGru,SchmuGRG,SchmuInt1,SchmuInt2}, respectively.

In the following part of this paper we develop the post-Newtonian
approximation of PUFT (field equations, equation of motion of a test
body and corresponding astrophysical applications). Our results
allow to compare the predictions of PUFT and the Einstein theory
with the experimental data.

\section{Four-dimensional field equations and
mechanical equations of motion (continuum and point-like test body)}

\subsection{Field equations in the space-time}

The 4-dimensional field equations of PUFT were received by
projecting the 5-dimensional field equations of the projective space onto 4-dimensional
space-time. In the following we present the 4-dimensional field
equations of PUFT in the Gauss system of units (Latin indices run
from 1 to 4, Greek indices from 1 to 3; the signature of the metric
is $(+,+,+,-)$; comma and semicolon denote partial and covariant
derivatives, respectively) \cite{Schmutzer1,Schmutzer2,SchmuGru}.
\subsubsection{Generalized gravitational field equation}
\begin{equation}\label{N1}
    R^{mn}-\frac{1}{2}\,g^{mn}R
    -{\Lambda_{\mathrm{S}}}\,\mathrm{e}^{-2\sigma}g^{mn}=
    \kappa_0(E^{mn}+S^{mn}+\Theta^{mn})\,.
\end{equation}
Here
\begin{align}\label{NN1}
R_{mn}=R^i\° _{mni}  \quad\text{and} \quad R=R^m\° _m
\end{align}
      are the usual
4-dimensional curvature quantities. Further one should realize the
cosmological term on the left-hand side of the field equation, being
the analog to that of the Einstein theory (General Relativity), but
here the scalaric field $\sigma$ is involved. Further
$\Lambda_{\mathrm{S}}$ is the  scalaric-cosmological
constant in PUFT with physical dimension of reciprocal square of  length.

In the basic field equations of Schmutzer's 5-dimensional PUFT a cosmological function is present, instead of the cosmological constant  of Einstein's 4-dimensional field theory. Here we should mention that this situation is similar to the ansatzes  of the quintessence in the cosmological models being intensively investigated at present time.

Let us in this context point to the relation
\begin{equation}\label{N1a}
    \kappa_0=\frac{8\pi\GN
    }{c^4}
\end{equation}
used by Schmutzer under following aspects: Subject to eventual later
theories with this Einstein gravitational constant
$\kappa_0$, as well  the Newtonian gravitational constant $\gamma_{\mathrm{N}}$ and the vacuum velocity of light $c$ are true   constants of
Nature.  With
respect to the numerical values, of course, one has to be careful,
since according to the precise measurements of the last decade it
seems that the measured ``Newton gravitational parameter'' $G$ is a
time-dependent quantity. Our subsequent treatment of the
post-Newtonian approximation shows a conceivable relationship
between these two quantities  $\GN$ and $G$. In the
present time the numerical values of both are closely neighboring.

The energy-momentum tensor of the non-geometrized matter (substrate)
is denoted by $\Theta^{mn}$. Further the following notations are
used:

\settowidth{\temp}{(energy-momentum tensor of}

\begin{align}\label{N2}
&\abc{a}E^{mn}=\frac{1}{4\pi}\,\Bigl(B^{mk}H_k\°
^n+\frac{1}{4}\,g^{mn}B_{kl}H^{kl}\Bigr)&&\parbox[t]{\temp}{(electromagnetic energy-momentum tensor),}
\notag \\[1ex]
&\abc{b}S^{mn}=\frac{2}{\kappa_0}\,\Bigl(\sigma^{\Sp m}\sigma^{\Sp
n}-\frac{1}{2}\,g^{mn}\sigma_{\Sp k}\sigma^{\Sp
k}\Bigr)&&\parbox[t]{\temp}{(scalaric energy-momentum tensor).}
\end{align}
The electromagnetic field strength tensor $B_{ij}$     and the
electromagnetic induction tensor  $H_{ij}$    will be explained in
context with the electromagnetic field equations.

As Schmutzer earlier pointed out, in  the electromagnetic
situation it is physically suggestive (without changing the
physical content) to absorb the occurring factor on the right-hand
side of the equation (\ref{N2}b) by introducing instead of the
dimensionless scalaric field function $\sigma$ the modified scalaric
field function (with a corresponding physical dimension)
$\Hat{\sigma}=\sqrt{\dfrac{2}{\kappa_0}}\,\sigma$. But with respect
to the approximation procedure we will keep to the more convenient
$\sigma$.

Specialization to a  perfect fluid gives in concretization of the energy-momentum tensor
$\Theta^{mn}$ the result
\begin{equation}\label{N3}
\Theta^{mn}=-\Bigl(\mu+\frac{p}{c^2}\Bigr)u^mu^n-p\,g^{mn}\quad\parbox[t]{1.32\temp}{($u^m$
four-velocity, $\mu$ mass density,  $p$   pressure).}
\end{equation}
(The minus sign is connected with the definition \eqref{NN1} of Ricci tensor.)
\subsubsection{Generalized electromagnetic field equations}
These basic equations read as follows:
\begin{align}\label{N4}
    &\abc{a} H^{mn}\° _{\Sr n}=\frac{4\pi}{c}\,j^m &&\text{(inhomogeneous
    system),}\notag\\
    &\abc{b} B_{<ij\Sp k>}=0&&\text{(cyclic system),}\notag\\
    &\abc{c}H^{mn}=\varepsilon B^{mn}\quad\text{with}\notag\\
    &\abc{d}\varepsilon=\mathrm{e}^{2\sigma}&&\text{(vacuum dielectricity/polarisation).
    }
\end{align}
The quantity  $j^m$    means the electric current density, e.g. in
the convective case: $j^m=\varrho u^m$, where $\varrho$ is the
 charge density.

\subsubsection{Scalaric field equation}
This equation has the form:
\begin{equation}\label{N5a}
    \sigma^{\Sp k}\° _{\Sr
    k}-\Lambda_{\mathrm{S}}\,\mathrm{e}^{-2\sigma}=-\frac{\kappa_0}{2}\,\Bigl(\frac{1}{8\pi}\,B_{ij}H^{ij}+\vartheta
    \Bigr)\,.
\end{equation}

In this equation  the scalaric substrate energy density $\vartheta$
(short: scalerg density) occurs,
which is a basically new quantity within the framework of the
traditional 4-dimensional physics (see \cite{Schmutzer2,SchmuGru}). Later we shall try to interpret
it.

\subsubsection{Equation of motion of the electrofluid and electric continuity equation}
Analogously to the procedure in the Einstein theory, by covariant
differentiation of the above field equations (\ref{N1}) and
(\ref{N4}a) we receive the following balance equations:
\begin{equation} \label{N6a}
   \abc{a}       \Theta^{mn}\° _{\Sr n} = -\frac{1}{c}\,B^{m}\° _{k}j^{k} +
          \vartheta\sigma^{\Sp m}\,,\quad\abc{b}j^m\° _{\Sr m}=0\,.
\end{equation}

For a perfect electrofluid by means of (\ref{N3})  the
corresponding equation of motion results:

\begin{align}\label{N6b}
    \Bigl(\mu+\frac{p}{c^2}\Bigr)u^m\° _{\Sr
    k}u^k=\frac{\varrho}{c}\, B^m\° _k u^k-\Bigl(p^{\Sp
    m}+\frac{1}{c^2}\,\deriv{p}{\tau}\,u^m\Bigr)\notag\\-
    \vartheta\Bigl(\sigma^{\Sp
    m}+\frac{1}{c^2}\,\deriv{\sigma}{\tau}\,u^m\Bigr)\,.
\end{align}

\subsection{The equations of motion of a point-like test body}
As we realized, the equation of motion in PUFT (as in the Einstein
theory) can be  derived  from  the  field  equations.  Of course, its
explicit form  requires  concrete  assumption  for  the
energy--momentum tensor $\Theta^{mn}$, electric four-current density
$j^{k}$ and the scalerg density $\vartheta$.

As it is well-known, in the Einstein theory for incoherent  matter
$\Theta^{mn} = - \mu u^mu^n $ and $j^m = \rho u^m$  ($\mu$  mass
density, $\rho$ electric charge density). This  means  for  the
case of a point-like test body (material point) with the  inertial
mass $M$ and  charge  $Q$ that  the following formulas are valid:
\begin{align} \label{N7}
 &\Theta^{mn} = -c\int \frac{M \delta^{4}(x-\xi(\tau))u^{m}u^{n}}{
 \sqrt{g(\xi(\tau))}}\dd\tau ,
\\[1ex] &
 \label{N8}
 j^{m} = Qc\int \frac{\delta^{4}(x-\xi(\tau))u^{m}}{
 \sqrt{g(\xi(\tau))}}\dd\tau  \,,
\end{align}
where the integrals in the equations  (\ref{N7})  and (\ref{N8})
must be taken along the world  line  (determined by  $\ds  x^i  =
\xi^i(\tau) $) of the test body; $\ds u^m =  \deriv{\xi^m}{\tau}$ is
the 4-velocity of the test body and $g =  -\det(g_{ij})$. By
substituting \eqref{N7} and \eqref{N8} into equation (\ref{N6a}) we
obtain the condition
\begin{equation}\label{N9}
        \vartheta\sigma^{,i} = -c\int
        \frac{\delta^{4}(x-\xi(\tau))}{
 \sqrt{g(\xi(\tau))}}\Biggl[\Deriv{(Mu^i)}{\tau} -
       \frac{e}{c}\,
       {B^i}_ku^k\Biggr]\dd\tau\, ,
\end{equation}
from wich the following structure of $\vartheta$ follows directly:
\begin{align}\label{N10}
 \vartheta= c^{3}\int \frac{{\cal M}\delta^{4}(x-\xi(\tau))}{
 \sqrt{g(\xi(\tau))}}\,\dd\tau \,.
\end{align}
The multiplier $c^3$ in  the  expression  \eqref{N10} guarantees
that the new introduced quantity ${\cal M}$, which is called
``scalaric mass'' (scalmass), has the same physical dimension as the
inertial mass $M$. Therefore the quantity $D  = {\cal M}c^2$ is
named ``scalaric substrate energy'' (short: scalerg
\cite{Schmutzer1}).

By  substituting  \eqref{N7}, \eqref{N8}
and \eqref{N10}   into equations (\ref{N6a}) we obtain
\begin{eqnarray} \label{N11}
          \Deriv{(M u^{i})}{\tau} =
          \frac{Q}{c}\,B^{i}\° _{k}u^{k} -
          c^{2}{\cal M} \sigma^{,i}\,.
\end{eqnarray}
Multiplying (\ref{N11}) by $u_i$ and keeping in mind that  $u^i u_i
= -  c^2$ ,  instead  of  (\ref{N11})  we find  the  following
equations which describe the motion of a point-like test body in a
gravitational, electromagnetic and scalaric field:
\begin{equation}    \label{N12}
          M\Deriv{u^{i}}{\tau}=\frac{e}{c}\,B^{i}\° _{k}u^{k}-
          c^{2}{\cal M}P^i\° _j \sigma^{,j}\,,
\end{equation}
\begin{equation}
          \deriv{M}{\tau}={\cal M}\sigma^{,k}u_{k}
          ={\cal M}\deriv{\sigma}{\tau}\,,
\label{N13}
\end{equation}
where $ P^i\° _j $ is the projection tensor,
\begin{align}\label{N13a}
    P^i\,_k = g^i\,_k + \frac{1}{c^2}u^iu_k\,, \quad P^i\,_ku^k  =
    0\,.
   \end{align}
We mention that the equations (\ref{N12}) and (\ref{N13})  coincide
with the corresponding ones performing  a transition from  the
equation  of motion  of  a perfect  fluid \cite{Schmutzer1}. From
the equation (\ref{N13}) follows that there exists a dependence of
the inertial mass (with particular features)  of the test body in
PUFT. Further one realizes that in the case of  a vanishing scalaric
field PUFT goes over into the  Einstein  theory \cite{Schmutzer1} in
which $M=M_{0}= \mathrm{const}$. Therefore  as a consequence of
(\ref{N13}) it is quite natural to suppose that the variability of
the inertial mass is exclusively caused  by  the scalaric field,
i.e. $M=M(\sigma)$. Hence follows that the scalaric mass  is
determined, in correspondence with equation (\ref{N13}), by $  \ds
{\cal M}=\deriv{M(\sigma)}{\sigma}$. In the case of an electrically
neutral particle ($ Q = 0$)  or  in the absence of an external
electromagnetic  field  ($B_{mn}=0 $) equation (\ref{N12}) reads:
\begin{equation} \label{N14}
         \Deriv{u^{i}}{\tau}=
          -c^{2}\eta (\sigma)\left(\sigma^{,i}+\frac{1}{c^{2}}\,\sigma^{,k}u_{k}u^{i} \right)\,,
\end{equation}
where  $\ds  \eta(\sigma)=\frac{{\cal  M}(\sigma)}{M(\sigma)}$.    The equation
(\ref{N14}) according to the concept of Schmutzer \cite{Schmutzer1}
contains an arbitrary function $\eta(\sigma)$ .

Let us note that we can give another  interpretation of the both masses $M$ and $\mathcal{M}$. In particular we write
\begin{align}\label{NN2}
    M(\sigma)=M_0f(\sigma)\quad\text{and}\quad \mathcal{M}=M_0 f'(\sigma)\quad(M_0=\mathrm{const})\,,
\end{align}
where $f(\sigma)$ as yet  is an arbitrary function of the $\sigma$-field.
Further we will call the constant $M_0$ inertial  mass of the  point-like body.
Thus as well as in the general relativity  the inertial mass remains a constant, but the equations of motion change:
\begin{align}\label{NN3}
    M_0\Deriv{u^i}{\tau}=-c^2\mathcal{M}\,P^i\°_{j}\sigma^{\Sp j}\quad
    \Bigl(\mathcal{M}=M_0\,\frac{f'(\sigma)}{f(\sigma)}\Bigr)\,.
\end{align}
Obviously, both approaches are mathematically  equivalent. But the second approach (see \eqref{NN3}) is  preferable from the physical point of view.

Let's note that equations of motion \eqref{NN3} can be rewritten in the form of Lagrange equations
\begin{align}\label{NN4}
    \deriv{}{\tau}\left(\pderiv{\mathcal{L}}{\dot{x}{}^{i}}\right)-\pderiv{\mathcal{L}}{x^i}=0\,,
    \quad \dot{x}{}^{i}=\deriv{x^i}{\tau}
\end{align}
with
\begin{align}\label{NN5}
    \mathcal{L}=-M_0cf(\sigma)\sqrt{-g_{ij}\dot{x}{}^{i}\dot{x}{}^{j}}
    +\frac{Q}{c}\,A_i\dot{x}{}^{i}\,,
\end{align}
where $A_i$ is the electromagnetic 4-potential. For the electrically neutral bodies ($Q=0$) the equations of motion   are equivalent to the equations of time-like geodesics in the pseudo-Riemannian space with the conformally transformed metric
\begin{align}\label{NN6}
g_{ij}\quad\rightarrow\quad\tilde{g}_{ij}\,,\quad\text{where}\quad\tilde{g}_{ij}=f^2(\sigma)\,g_{ij}\,.
\end{align}

Let as mention that from (\ref{NN3}) one immediately learns that the
weak equivalence principle is fulfilled exactly if $f(\sigma)$ is a
universal function for all kind of matter.
Obviously, given relation for extended bodies can  be fulfilled only approximately as $M$ and $\mathcal{M}$ have the different physical nature.

The
post-Newtonian approximation of PUFT for the case
\begin{align}\label{N15a}
  M(\sigma)=M_{0}e^{\eta_0\sigma}, \quad
         {\cal M}(\sigma)= (\eta_0 M_0)\mathrm{e}^{\eta_0 \sigma}\,,
\quad (\eta_0=\mathrm{const})
\end{align}
 was
constructed in the paper \cite{Gor2}.
In this context we note that just the dependence between the
inertial and scalaric masses in the form
\begin{align}\label{N15b}
    M(\sigma)=\mathcal{M}\sigma, \quad \mathcal{
M}(\sigma)=\mathrm{const}\,,
\end{align}
and  some other dependencies
 were studied and elaborated in detail by
Schmutzer for various cosmological models. The results of this
cosmological investigations are resumed in the monographs \cite{Schmutzer3,SchmuGru}.

\section{Post-Newtonian  Approximation of the PUFT}\label{sec3}

\subsection{Introduction}

Let us consider a system of slowly moving bodies bounded by
gravitational interaction (e.g. planetary system). In order to
describe the gravitational field of such a system at large distances
from its center, in the Einstein theory one can use the so-called
post-Newtonian approximation. Here we will show that the field
equations of PUFT also allows an analogous approximation. Similar to
the procedure in the Einstein theory it is convenient to take the
ratio $\beta = \ds v/c$ as a small expansion parameter of the exact
field equations, where $v$ is the characteristic velocity of the
motion of the bodies, which is related to the gravitational
potential $\phi$ as follows:
\begin{equation}  \label{N16}
          {v^{2}\over c^2 }\sim {\phi\over c^2 }\sim \beta^{2}
\end{equation}
(in the planetary system  $\beta \sim 10^{-4}$ to $10^{-3}$).

For the further investigation we restrict our considerations to the
following assumptions:
\begin{enumerate}
  \item
  The electromagnetic field is equal to zero: $B^{ij}=0$.
  \item
  According to the suppositions \eqref{NN2} and \eqref{NN3} we assume that
\begin{equation}\label{N17}
  \frac{f'(\sigma)}{f(\sigma)}\ll 1
\end{equation}
for all bodies.
  \item  The energy--momentum tensor of the perfect fluid has the
  standard form \eqref{N3}
\item
For the system considered the equations \eqref{N5a} are joint with
the assumption that the scalaric field may be introduced as a
superposition of two fields
\begin{equation}\label{N18}
  \sigma=\si+\so\quad(|\si|\ll|\so|)\,,
\end{equation}
whose first one ($\si$) has sources inside the considered system of
bodies, and the second one ($\so$) refers to the outside of it. In particular,
$\so{}$ may have a cosmological origin. Thus by consideration of the
motion of bodies inside the planetary system we may regard the field
$\so{}$ as a quasi constant field: $(\so{})_{\Sp k}\approx 0$.
\item
We suppose that there exists a coordinate system, in which in zeroth
order approximation the metric tensor equals the Minkowski tensor
$\eta_{ij} \equiv {\rm diag}(1,1,1,-1)$. Then the following power
series approximation is possible:
    \begin{subequations}\label{N19}
\begin{align} \label{N19a}
          g_{\alpha\beta} & =  \delta_{\alpha\beta} +
{} \stackrel{2}{g}\° _{\alpha\beta}  + O(\beta^4)\,, \\
     \label{N19b}
          g_{\alpha 4} & ={} \stackrel{3}{g}\° _{\alpha 4} + O(\beta^5)\,, \\
     \label{N19c}
          g_{44} & =  - 1 +{} \stackrel{2}{g}\° _{44}+\stackrel{4}{g}\° _{44}
          + O(\beta^6)\,;
\end{align}
\end{subequations}
and for the stress tensor $T^{mn}\equiv \Theta^{mn}+S^{mn}+E^{mn}$:
\begin{subequations}\label{N20}
\begin{align} \label{N20a}
          T^{\alpha\beta} & ={} \stackrel{0}{T}\° ^{\alpha\beta} +
{} \stackrel{2}{T}\° ^{\alpha\beta}  + O(\beta^4)\,, \\
     \label{N20b}
          T^{\alpha 4} & ={} \stackrel{1}{T}\° ^{\alpha 4} + O(\beta^5)\,, \\
     \label{N20c}
          T^{44} & ={} \stackrel{0}{T}\° ^{44} +{} \stackrel{2}{T}\° ^{44}+\stackrel{4}{T}\° ^{44}
          + O(\beta^6)\,.
\end{align}
\end{subequations}

\end{enumerate}

\subsection{Newtonian-like approximation of PUFT}

In the case of a vanishing electromagnetic field the  equations
(\ref{N1})  and  (\ref{N5a})  in consideration of \eqref{N18} lead  to  the Newtonian-like approximation:
\begin{equation} \label{N21}
          \abc{a}   \triangle\phi=4\pi \GN \mu\,, \qquad
          \abc{b}   \triangle\si=-\frac{4\pi\GN}{c^2}\,\mut\,,
\end{equation}
where $\phi$ is the  gravitational  potential,  which is connected
with the metric:
\begin{align}\label{N21a}
 \ds g_{44} = - 1 - \frac{2\phi}{c^2}\,;
\end{align}
$\mu=-\dfrac{1}{c^2}\stackrel{0}{T}{}^{44}$ is the mass density, and
$\mut=\dfrac{1}{c^2}\,\vartheta$ is the so-called scalaric mass
density (scalmass density). In the case of $N$ gravitationally
bounded point-like bodies the quantities $\mu$ and $\mut$ in the
Newtonian-like approximation take the form ($A$ running from 1 to $N$)
\begin{align}\label{N22}
    \abc{a}\mu=
    \sum_A M_A\,\delta^{(3)}(\bm{r}-\bm{r}_A)\,,\quad
    \abc{b}\mut{}=
    \sum_A \mathcal{M}_A\,\delta^{(3)}(\bm{r}-\bm{r}_A)\,.
\end{align}
The solution of the equations \eqref{N21} which vanishes at infinity
can be written as
\begin{align}\label{N23}
    \abc{a}\phi(\bm{r}_K)=-\GN\sum_{A\neq
    K}\frac{M_A}{r_{AK}}\,,
    \quad\abc{b}
    \si(\bm{r}_K)=\GN\frac{1}{c^2}\sum_{A\neq
    K}\frac{\mathcal{M}_A}{r_{AK}}\,,
\end{align}
where the abbreviations
\begin{align}\label{N24}
    r_{AK}=\left|\bm{r}_{AK}\right|\,,\quad
    \bm{r}_{AK}=\bm{r}_A-\bm{r}_K
\end{align}
are used.

Taking into account the assumption \eqref{N18} and the solutions
\eqref{N23}, we obtain the following equations of motion for $N$
gravitationally bounded bodies in the Newtonian-like approximation of
PUFT:
\begin{align}\label{N25}
    \deriv{\bm{v}_K}{t}=-\GN\sum_{A\neq
    K}M_A\Bigl(1-\frac{\mathcal{M}_A}{M_A}\,\frac{\mathcal{M}_K}{M_K}\Bigr)
    \frac{\bm{r}_K-\bm{r}_A}{\left|\bm{r}_K-\bm{r}_A\right|^3}\,.
\end{align}
Let us notice that up to now both  the inertial mass $M$ and the
scalaric mass $\mathcal{M}$ are in general completely independent.
However for fulfilling the weak equivalence principle   it is
necessary that the ratio of both  masses  is for all bodies the
same universal function of the scalaric field.
As it was mentioned above, the equivalence principle
has already been tested experimentally to a
relative precision of $10^{-12}\div 10^{-14}$ \cite{Will}. Thus the function  $f(\sigma)$ (see \eqref{NN2}) at least in the Sun-system is in very good approximation an universal function: $\dfrac{|f_{K_1}-f_{K_2}|}{f_{K_1}}<10^{-12}$.
Taking into account the assumptions \eqref{N18} and \eqref{N17}
we can write for all bodies
\begin{align}\label{N26a}
    M_K(\sigma)=f(\sigma)M_{0K}\,,\quad
    f(\sigma)\approx f(\so)\left(1+\frac{f'(\so)}{f(\so)}\si\right)\,.
\end{align}
With the help of the last relation we can put the equation of
motion into the Newtonian-like form
\begin{align}\label{N26}
     \deriv{\bm{v_K}}{t}=-G_{\mathrm{S}} \sum_{A\neq
    K}\,M_A
    \frac{\bm{r}_K-\bm{r}_A}{\left|\bm{r}_K-\bm{r}_A\right|^3}=
    -G_{\mathrm{S}} \sum_{A\neq
    K}\,M_{0A}f(\so)\,
    \frac{\bm{r}_K-\bm{r}_A}{\left|\bm{r}_K-\bm{r}_A\right|^3}\,,
\end{align}
where we have introduced the   ``scalaric-gravitational
parameter'' $G_{\mathrm{S}}$ (see \cite{Schmutzer2,SchmuGru}):
\begin{align}\label{N27}
    G_{\mathrm{S}}\equiv\GN\bigl(1-\sig^2\bigr)\approx 6.6726\times
10^{-8}\,\mathrm{\dfrac{cm^3}{g\;
  s^2}}\,,
\end{align}
which   numerically coincide at the present time with the conventional Newtonian gravitational constant.
Here
\begin{align}\label{NN6a}
    \sig=\frac{f'(\so)}{f(\so)}=\frac{\mathcal{M}(\so)}{M(\so)}\simeq\mathrm{const}\quad
    (\so=\so(t))\,,
    \quad\sig\ll 1\,.
\end{align}

Summarizing the last results, we  can write the field equations
\eqref{N21} and the equation of motion of a test point-like body in
the Newtonian-like approximation of PUFT in the familiar form:
\begin{equation} \label{N28}
          \abc{a}   \deriv{\bm{v}}{t}=-\nabla\PHI\,,\qquad
          \abc{b}   \triangle\PHI=4\pi G_{\mathrm{S}}\mu\,,
\end{equation}
using the quantity
\begin{align} \label{N29}
\PHI=\phi+c^2\frac{f'(\so)}{f(\so)}\, \si=\phi+c^2\sig\, \si
\end{align}
as effective (empirical) Newtonian potential.

From these considerations we learn that the scalaric-gravitational parameter $G_{\mathrm{S}}$ is not a constant in strict sense, since
it depends on the scalaric field $\so$ which was introduced  for the description of the effective   cosmological influence. Therefore $\dot{G_{\mathrm{S}}}\equiv\deriv{G_{\mathrm{S}}}{t}\neq 0$, but $\dfrac{\dot{G_{\mathrm{S}}}}{G_{\mathrm{S}}}$ is very small. Because of this extreme
smallness up to now direct measurements of $\dfrac{\dot{G_{\mathrm{S}}}}{G_{\mathrm{S}}}$ in
laboratories are without success. But nevertheless for the
Earth-Moon system indirect estimates of $\dfrac{\dot{G_{\mathrm{S}}}}{G_{\mathrm{S}}}$,
received as result of evaluating the motion of the lunar orbit, give
magnitudes till  $10^{-14}\div10^{-13}\kk{s^{-1}}$, particularly
E.\,V.\,Pitjeva: $10^{-14}\kk{s^{-1}}$ \cite{Pitjeva}. However one
should remember that these numerical results are received within the
framework of Newtonian celestial mechanics, being corrected by
admitting a time-depending gravitational constant. Of course, in the
case of PUFT the evaluation of the same observed data can yield
essentially differing results, being of high importance for the
numerical proof of PUFT.

Let us in this context emphasize that the equation (\ref{N21}b) was
obtained for the case of vanishing external electromagnetic fields
from the basic equation (\ref{N5a}), in which as source term for the
scalaric field also the (electromagnetic) Larmor invariant
$\Lambda_{\mathrm{L}}$=$B_{ij}H^{ij}$ occurs. In the previous
calculations this term has been neglected. But for deeper
understanding of the electromagnetic influence on scalarism the
Larmor term has to be taken into account. For example, this term
should not be ignored for a detailed treatment of the inner region
of electrically neutral matter, where local electromagnetic fields
of appreciable strength occur, inducing local scalaric fields.
According to (\ref{N5a}) these local scalaric fields
$\sigma_{_{loc}}$  may yield a considerable contribution to the
global  $\sigma$-field appearing in (\ref{N21}b).

Let us further mention that recently a review article on various 4-dimensional approaches of theories with time-dependent cosmological gravitational parameters appeared, comparing the ansatzes with corresponding measuring results (see \cite{Sanders}).

Concluding this subsection, let us draw our attention to another
interesting subject. As it is well known, the first measurement of
the frequency red shift in a gravitational field (experiments of
Pound and Rebka \cite{Pound1} and Pound and Snider \cite{Pound2})
shows that the free motion of bodies corresponds to geodesics up to
very high precision.   Looking at \eqref{N14}  one immediately
recognizes that in the PUFT the free fall is not the geodesic
motion. This fact means that PUFT is a non-metric gravitational
theory in the sense of this definition. In the Newtonian-like
approximation we can say that the red shift is determined by the
metric, i.e. by the gravitational potential $\phi$ (see
\eqref{N21a})
\begin{align}\label{N30}
     \frac{\Delta \nu}{\nu}=-\frac{\Delta\phi}{c^2}=
     -\bigl(1-\sig{}^2\bigr)^{-1}\frac{\Delta\PHI}{c^2}\,.
\end{align}
In contrast to this result the gravitational force depends on the
effective Newtonian potential $\PHI$. In the conventional notation
the red shift formula has the shape
\begin{align}\label{N31}
    \frac{\Delta \nu}{\nu}=(1+\alpha)\frac{\Delta
   U}{c^2}\quad\text{($U=-\PHI$)}\,,
\end{align}
Hence we find that $\alpha\approx \sig^2$.

Numerous experiments were performed to check the formula \eqref{N31}
(\cite{Will1}). Without further discussion of the experimental data
obtained, we notice that a reliable upper bound of $\alpha$ is given
by \cite{Will}: $|\alpha|<2\pot{-4}$. Hence we obtain a first
estimate for $\so$ at the present time,  following from the red
shift experiments:
\begin{align}\label{N32}
   \sig^2<2\pot{-4}\quad\text{or}\quad \frac{f'(\so)}{f(\so)}=\sig<1.4\pot{-2}\,.
\end{align}

\subsection{The first post-Newtonian approximation}
Using the solution of the field equations in Newtonian-like approximation
(see the next section) we claim that the expansion of the $\si $
starts with a term of the order $ \sim \beta^2$:
\begin{eqnarray}    \label{N33}
          \sigma    = \so   +\frac{1}{c^2}\Lambda + \frac{1}{c^4}\lambda
          +O(\frac{1}{c^6}).
\end{eqnarray}

Substituting the expansion series (\ref{N19}), (\ref{N20}) and
\eqref{N33} in the field equations and taking into account the
explicit expression (\ref{N3}) for the stress-energy tensor, we
receive a set of equations, which simply may be integrated in
so-called harmonic coordinates defined by $\Gamma^i\° _{jm}g^{jm} =
0$. For simplicity omitting the intermediate results, we finally
arrive at the outcome:
\begin{subequations}\label{A}
\begin{align} \label{Aa}
&
          g_{\alpha \beta}  =  \delta_{\alpha \beta}
          \left(\ds 1 - \frac{2\phi}{c^2} \right) + O(\beta^4),\\
\label{Ab}
 &         g_{\alpha 4}  =   \frac{1}{c^3}\,\xi_{\alpha}
          + O(\beta^5),\\
          \label{Ac}
&          g_{4 4}  =
          - \left[1 + \frac{2 \phi}{c^2} + \frac{2}{c^4}
          \left(\phi^2 + \chi \right) \right] +
           O(\beta^6),
\end{align}
\end{subequations}
where the abbreviations
\begin{equation}  \label{D.1}
          \phi({\bm x},t)=-\GN\int{{\mu ({\bm x}^{'},t)
          \over |{\bm x}-{\bm x}^{'}|}d^{3}{\bm x}^{'}},
\end{equation}
\begin{equation} \label{D.2}
          \xi_{\alpha}({\bm x},t)=-
          4\GN\int{{\mu({\bm x}^{'},t) v_{\alpha}({\bm x}^{'},t)\over
          |{\bm x}-{\bm x}^{'}|}d^{3}{\bm x}^{'}},
\end{equation}
\begin{equation}\label{n6}
  \chi=\Phi_3-2\Phi_2+2\Phi_1+3\Phi_4,
\end{equation}
\begin{equation} \label{D.4}
          \Phi_{1}({\bm x},t)=-\GN\int{{\mu({\bm x}^{'},t) v^{2}({\bm
x}^{'},t)
          \over |{\bm x}-{\bm x}^{'}|}d^{3}{\bm x}^{'}},
\end{equation}
\begin{equation} \label{D.5}
          \Phi_{2}({\bm x},t)=-\GN\int{{\mu({\bm x}^{'},t) \phi({\bm
x}^{'},t)
          \over |{\bm x}-{\bm x}^{'}|}d^{3}{\bm x}^{'}},
\end{equation}
\begin{equation}\label{n7}
  \Phi_3({\bm x},t)=-{1\over 4\pi}\int{\pderiv{{}^2\phi({\bm
x}^{'},t)}{t^{2}}
          {1\over |{\bm x}-{\bm x}^{'}|}d^{3}{\bm x}^{'}},
\end{equation}
\begin{equation} \label{D.6}
          \Phi_{4}({\bm x},t)=-
          \GN\int{{p({\bm x}^{'},t)\over |{\bm x}-{\bm x}^{'}|}d^{3}{\bm
          x}^{'}},
\end{equation}
\begin{align}\label{n10}
  \vartheta={c^2\mu}{\sig }\Bigl[1+
  \frac{\phi}{c^2}\sig^2+\frac{1}{c^4}\,(\sig^4\phi^2-
  \lambda\sig)\ +O(1/c^6)\Bigr]\,,
\end{align}
\begin{equation}\label{n8}
  \sigma=\so \Bigl\{1 - \frac{\sig^2}{c^2}\,\phi-
  \frac{\sig^2}{ c^4}\left[\chi-2\Phi_1-3\Phi_4\right]+O(1/c^6)\Bigr\}\,,
\end{equation}
were used. Here $\mu$ and $p$ are mass density and pressure respectively.

In the preceding relations, as well as in the subsequent equations
of motion for a neutral test particle, the terms of the order of
magnitude $\ds \frac{v^2}{c^2} \, {\sig  ^4}$ were neglected:
\begin{align} \label{4.17}\notag
          \deriv{{\bm v}}{t} &=
          -\bigl(1-\sig^2\bigr){\bm \nabla}
          \left(\phi+{2\over c^2}\,\phi^{2}+{1\over c^2}\,\chi\right)\\
          &-
         \bigl(1+\sig^2\bigr) {v^{2}\over c^2}\,{\bm \nabla}\phi
    + 3{{\bm v}\over c^2}
          \pderiv{\phi}{t}\bigl(1+{3\sig ^2}\bigr)
\notag\\ &
          -{1\over c^2}\pderiv{{\bm \xi}}{t}+
         {{\bm v}\over c^2}\times({\bm \nabla}\times{\bm \xi})+
         {4{\bm v}\over c^2}\,({\bm v}{\bm \nabla})\phi\notag
      \\
         &- \frac{\sig ^2}{ c^2}\,{\bm \nabla}\left(2\Phi_{1}+3\Phi_{4}
         \right) + O(\beta^4).
\end{align}

\subsection{The equations of motion of  ${N}$ gravitationally
bounded point-like bodies in the first post-Newtonian approximation}

For the first time the approximated equations of motion within the
framework of the Einstein theory  were derived by  A. Einstein, L.
Infeld and B. Hoffmann (1938) as well later by a different approach
by V. Fock and N. M. Petrova (1939, 1949). These publications
initiated the start of a lot of papers which mainly aimed at the
derivation of the equations of motion from the field equations.

As already mentioned above, the equations of motion can be obtained
from the field equations of PUFT in the same way as in the Einstein
theory. Omitting intermediate results, now for the $N$ point-like
bodies $\bigl($compare with \eqref{N7} and \eqref{N8}$\bigr)$ we present the
substrate stress tensor $\Theta^{mn}$ and the scalmass density $\mut
$:
\begin{align}\label{N34}
    &\Theta^{mn}(\bm{r})=-\sum_{A=1}^{A=N} M_A\frac{1}{\sqrt{-g}}
    \deriv{x_A^m}{t}\,\deriv{x_A^n}{t}\,\deriv{t}{\tau}\,\delta^{(3)}(\bm{r}-\bm{r}_A)\,,\\
    &\mut \equiv{}\frac{1}{c^2}\,\vartheta=
    \sum_{A=1}^{A=N}\mathcal{M}_A\label{N34a}
    \frac{1}{\sqrt{-g}}\,\deriv{t}{\tau}\,\delta^{(3)}(\bm{r}-\bm{r}_A)\,.
\end{align}
Taking into account the explicit expression \eqref{A} for the
metric, then we find in the first post-Newtonian approximation the
following expressions for the mass density and the scalmass density:
\begin{align}\label{E1}
    \mu(\bm{r})=\sum_{A}M_A\Bigl(1-\frac{v^2_A}{2c^2}+\frac{3\phi}{c^2}\Bigr)\delta^{(3)}(\bm{r}-\bm{r_A})+O(\beta^4)\,,
\end{align}
\begin{align}\label{E2}
    \mut(\bm{r})=
    \sum_{A}\mathcal{M}_A\Bigl(1-\frac{v^2_A}{2c^2}+\frac{3\phi}{c^2}\Bigr)\delta^{(3)}(\bm{r}-\bm{r_A})+O(\beta^4)\,.
\end{align}
Hence in this case the equations (\ref{N6a}) lead  to the following
equations of motion for the gravitationally bounded $N$ point-like
bodies within the framework of PUFT:
\begin{align}\label{EB}\notag
    \deriv{\bm{v_K}}{t}&=\sum_{A\neq K}\frac{\GN M_A
    \bm{r}_{AK}}{r_{AK}^3}\biggl[\bigl(1-{{ \sig^2}}\bigr)
    \biggl(1-4\sum_{B\neq K}
    \frac{\GN M_B}{r_{BK}c^2}\\
    &-\sum_{C\neq
    A}\frac{\GN M_C}{r_{CA}c^2}
    \Bigl(1-\frac{\bm{r}_{AK}\cdot\bm{r}_{CA}}{2r_{CA}^2}\Bigr)
    -\frac{3}{2c^2}\Bigl(\frac{\bm{v_A\cdot\bm{r_{AK}}}}{r_{AK}}\Bigr)^2\biggr)
    \notag\\
    &+\bigl(1+{{ \sig^2}}\bigr)v_K^2/c^2+2v_A^2/c^2-4\bm{v}_K\cdot\bm{v}_A/c^2\biggr]
    \notag\\&
    +\frac{1}{2}\,\bigl(7+{{ \sig^2}}\bigr)\sum_{A\neq
    K}\sum_{C\neq A}\bm{r}_{CA}\frac{\GN^2
    M_AM_C}{r_{AK}r_{CA}^3c^2}
    \notag\\
    &-\sum_{A\neq
    K}\bigl(\bm{v}_A-\bm{v}_K\bigr)
    \frac{\GN
    M_A\bm{r}_{AK}\cdot\Bigl[\bigl(3+{ \sig^2}\bigr)\bm{v}_A-4\bm{v}_K\Bigr]}{c^2r_{AK}^3}\,,
\end{align}
with $\GN=\bigl(1+{\sig^2}\bigr)\,G_{\mathrm{S}}=\mathrm{const}$.

This system of equations of motion can be considered as the analog
(in PUFT) to the Einstein-Infeld-Hoffmann equations (in the Einstein
theory).

\subsection{Perihelion Motion of Mercury}

         In this section we investigate the motion of  a  test  body  (e.g.
         Mercury) around a central body (e.g.  sun)  which  for  simplicity
         will be considered as non-rotating and spherically  symmetric.  As
         before we suggest that the condition $\sig \ll  1$  is  fulfilled.
         Under these assumptions the integration  of  the  field  equations
         leads to
\begin{equation} \label{4.18}
     \begin{array}{lll}            \ds
     \mbox{a)}\qquad
          g_{\alpha\beta} & = & \delta_{\alpha\beta}{\ds (1+
          {2\GN M_{c}\over c^{2}R}})
          + O(\beta^4),  \\ \medskip
     \mbox{b)} \qquad
          g_{\alpha 4} & = & 0,   \\ \medskip
     \mbox{c)} \qquad
          g_{44} & = & -1+{\ds {2\GN M_{c}\over c^{2}R}}-
          {\ds {2\GN ^{2}M_{c}^{2}\over c^{4}R^{2}}}
          + O(\beta^6),
     \end{array}
\end{equation}
         where $M_{c}$ is a constant coinciding with the inertial  mass  of
         the central body if the scalaric field vanishes. Therefore in this
         approximation the metric has the same form as the  metric  in  the
         Einstein theory. We have to note that E.\,Schmutzer \cite{Schmutzer1,SchmuExact1}
         succeeded in finding the exact spherically symmetric solutions in implicit form. In the paper \cite{Nato} we find three parametric exact  spherically symmetric solution in explicit form. In this paper we find this solution in the harmonic coordinates, which were used by obtaining the post-Newtonian equations. This solution in the corresponding approximation is identically to \eqref{4.18}.

         The equation of motion of the test body reads:
\begin{eqnarray} \label{4.19}
          \deriv{{}^2{\bm   R}}{t^{2}} & \equiv &\deriv{{\bm v}}{t}
          = -{\GN M_{c}\over R^{3}}{\bm R}\Biggl[\left(1-
          \sig^{2}\right)
          \left(1-{4\GN M_{c}\over c^{2}R}\right)
     \nonumber \\ \bigskip
      &+&\left(1+\sig^{2}\right){v^{2}\over c^2}\Biggr]
          +  {4\GN M_{c}\over c^{2}R^{3}}{\bm   v}({\bm  v  }{\bm R})
          + O(\beta^4).
\end{eqnarray}
         In the Newtonian approximation the equation (\ref{4.19}) goes over
         into the equation of motion
\begin{equation} \label{4.21}
          \deriv{{}^2 \bm R}{t^{2}} \equiv \deriv{{\bm v}}{t}
          =-{\GN M_{c}\over R^{3}}{\bm R}
         \left(1-\sig^2\right).
\end{equation}
         Integration leads to the following set of equations:
\begin{equation}    \label{4.22c}
          R^{2}\deriv{\varphi}{t}=\left[\GN M_{c}\left(1-
          \sig^{2}\right)P\right]^{1/2},
\end{equation}
\begin{equation} \label{4.22d}
          {\bm   v}\equiv \deriv{{\bm R}}{t}=
          \left[{\GN M_{c}(1-\sig^{2})\over P}\right]^{1/2}\left[
          -{\bm e}_{x}\sin\varphi+
          {\bm e}_{y}(\varepsilon+\cos\varphi)\right],
\end{equation}
         where
\begin{equation} \label{4.22}
     \mbox{a)} \quad
         {\bm R}=R({\bm e}_{x}\cos\varphi+{\bm e}_{y}\sin\varphi),
        \qquad \mbox{b)} \quad
          R=P(1+\varepsilon\cos\varphi)^{-1}.
\end{equation}
         As it is well-known, this solution describes  the  motion  of  the
         test body in a plane (spanned by the basis vectors ${\bm e}_x$ and
         ${\bm e}_y$) orthogonal to the angular momentum.  We  remind  that
         $R$ and $\varphi$ are polar coordinates in the  plane  of  motion,
         $P$ and $\varepsilon$ (eccentricity) are  the  parameters  of  the
         ellipse.

         Applying the  method  of  successive  approximation  we  find  the
         following post-Newtonian solution:
\begin{equation} \label{4.23}
          {\bm v}= \left[{\GN M_{c}(1-\sig^{2})\over
          P}\right]^{1/2}\left[-{\bm e}_{x}\sin\varphi+ {\bm e}_y
          (\varepsilon+\cos\varphi)\right]+\delta{\bm  v},
\end{equation}
\begin{equation} \label{4.24}
          R^{2}\deriv{\varphi}{t}\equiv
          ({\bm R}\times {\bm v})_z =
          \left[\GN M_{c}(1-\sig^{2})P\right]^{1/2}(1+\delta h),
\end{equation}
         where
\begin{equation} \label{4.25}
          \delta h=-{4\GN M_{c}\over  c^{2}P}\varepsilon\cos\varphi
\end{equation}
         and
\begin{eqnarray} \label{4.26}
          \delta {\bm v} & = & \sqrt{\frac{\GN M_{c}(1 -\sig^2)}{P} }
          \left({\GN M_{c}\over
          c^{2}P}\right) \Biggl\{ {\bm e}_{x}
          \biggl[ \sin\varphi\left((3-\varepsilon^{2})-\sig^{2}(1+\varepsilon^{2})\right)
     \nonumber \\ \medskip
          & - &
          \varepsilon(3+\sig^{2})\varphi
          +    {1\over
2}\varepsilon(1-\sig^{2})\sin2\varphi\biggr]
     \nonumber \\ \medskip
          & - &
          {\bm e}_{y}\biggl[\cos\varphi(1+\varepsilon^{2})(3 -\sig^{2})
          + {\varepsilon\over
2}\left(1-\sig^{2}\right)\cos2\varphi
          \biggr]\Biggr\}.
\end{eqnarray}
         Using the identity
\begin{equation} \label{4.27}
          \deriv{}{\varphi}{1\over R}=
          - \left(R^{2}\deriv{\varphi}{t}\right)^{-1}
          \left({{\bm R}{\bm v}\over R}\right),
\end{equation}
         we get from ({\ref{4.26}) for the trajectory of the test body:
\begin{eqnarray} \label{4.28}
          {P\over R} & = &
          1+\varepsilon\cos\varphi+
          {\GN M_{c}\over c^{2}P}\biggl[{1\over 2}\varepsilon\cos\varphi
          (7+\sig^{2}) \nonumber \\ \medskip
          & + & \varepsilon(3+\sig^{2})\varphi
          \sin\varphi\biggr].
 \end{eqnarray}
The calculation of the perihelion motion leads to the result
\begin{equation} \label{perig}
          \delta \varphi \approx \frac{6\pi \GN M_{c}}{c^{2}P}
          \left(1 + \frac{1}{3 }\,
          \sig^2 \right)
\end{equation}
         for one revolution. This formula (\ref{perig}) is  identical  with
         the relation found by Schmutzer \cite{Schmutzer2} using  the  exact
         spherically  symmetric  solution.  If  $\sig  =  0$, then the relation  (\ref{perig})
         coincides with the corresponding result in the Einstein theory.

Finally to this section \ref{sec3} we draw the attention of the reader at a new  paper by E.\,Schmutzer on a set
of possibilities to couple scalarism ($\vartheta$) to the usual 4-dimensional physics, being investigated by him \cite{Schmu16}.

\section{Conclusion}

         In the present paper we investigated the equation of motion  of  a
         point-like test body in PUFT and  the  possibility  of  functional
         dependence of the inertial mass of  an  external  scalaric  field.
         Although the idea of variability  of  inertial  mass  is  not  new
         itself ( see e.\,g. \cite{Bekenstein}), in  PUFT  it  appears  quite
         natural.

         Let us remind that in PUFT the gravitational central mass  $M_{c}$
         appears in the exact spherically symmetric solution. At  the  same
         time the acceleration of test bodies in an external  gravitational
         field  is characterized by the effective Newtonian potential $\PHI$ (see \eqref{N28} and \eqref{N29}). Using the definition \eqref{N27} we obtain for the solar system that $ M_{c}=M_{\odot}$,
         since the mass of the sun has  to  be
         determined by its gravitational interaction ($\PHI=-G_{\mathrm{S}}M_{\odot}/r$).  Hence  follows  that
         the perihelion motion of Mercury reads
\begin{equation} \label{C.1}
          \delta \varphi \approx \frac{6\pi G_{\mathrm{S}} M_{\odot}}{c^{2}P}\left(
          1 + \frac{4}{3 }\,\sig^2 \right).
\end{equation}
         The Einstein effects, as light  deflection  and  photon  frequency
         shift, are determined  exclusively  by  the  space-time  geometry.
         Therefore the results in PUFT coincide with the corresponding ones
         in the  Einstein  theory  if  $M_{c}$  is  used.
         Expressing the mass $M_{c}$ by the experimentally determined value
         of the solar mass $M_{\odot}$, we obtain  the  following  formulas
         for the light deflection:
\begin{equation} \label{C.2}
          \triangle \chi = \frac{4G_{\mathrm{S}} }{R_{\odot}c^2}M_{\odot}\left(1 +
          \sig^2\right)\,,
\end{equation}
         where $R_{\odot}$ is the radius of the Sun. Of course  (\ref{C.1})
         and (\ref{C.2}) can be applied for calculation  of  post-Newtonian
         effects only if corrections connected with  $\sig^2$  are  greater
         than the post-post-Newtonian ones.
If we compare the two last expressions with the experiment (see \cite{Will}) then we find immediately that from \eqref{C.1} and \eqref{C.2} follows that $\sig<0.048$         and $\sig<10\pot{-2}$, respectively. Let's remind that from the red shift experiment follows (see \eqref{N32}): $\sig<1.4\pot{-2}$.
         \\[2ex]

\textbf{Acknowledgement}

We are very grateful to  professor M. Schneider (formerly
Technical University of Munich) for his intended  comparison of PUFT with measuring
astrophysics, that has motivated  us to this approximate treatment of PUFT.

 A.\,K.\,Gorbatsievich is very grateful to DAAD and FSU Jena (Germany) for
    financial support and hospitality.


\begin{thebibliography}{99}


\bibitem{Will}
C. M. Will,   \emph{Living Rev. Relativity} \textbf{9} (2006), 3
(www.livingreviews.org/lrr-2006-3); ArXiv gr-qc/0510072).
\bibitem{Schmutzer1}
 E.   Schmutzer,
    \textit{Fortschr. Phys.} \textbf{43}, (1995) 613.
\bibitem{Schmutzer2}
E. Schmutzer,  \emph{Projektive Einheitliche Feldtheorie mit Anwendungen
in Kosmologie und Astrophysik. Neues Weltbild ohne Urknall? (Mit
einem Anhang von A.\,K. {Gorbatsievich})} (Verlag Harri Deutsch GmbH, Frankfurt am Main, 2004).
\bibitem{SchmuGru}
E.~Schmutzer,  \emph{F\"unfdimensionale Physik} (Wissenschaftsverlag Th\"u\-ringen, Lange\-wiesen 2009).

\bibitem{SchmuGRG}
E. Schmutzer, \emph{Gen. Rel. Gravit.} \textbf{33} (2001) 843.
\bibitem{SchmuInt1}
E. Schmutzer, \emph{J. Mod. Phys. E} \textbf{16} (2007) 1181.
\bibitem{SchmuInt2}
E. Schmutzer, \emph{J. Mod. Phys. E} \textbf{18} (2009) 1903.
\bibitem{Gor1}A. K. Gorbatsievich,  \emph{Gen. Rel.  Grav.} \textbf{33}, (2001) 965.
\bibitem{Schmutzer3}
E.    Schmutzer, in: \emph{Proceedings of the 17th Erice Course
    of the International School of Cosmology and Gravitation,}  (Eds.
    P.G.~Bergmann and V.~de Sabbata,  Kluwer Academic Publishers,
    Dordrecht, 2002), p. 387
\bibitem{Blin}
A. A. Blinkouski, and  A. K. Gorbatsievich,
\emph{Gravitation and Cosmology,} \textbf{7} (2001) 286.
\bibitem{Schmuhab}
 E. Schmutzer,  Habilatationsschrift at the Friedrich Schiller University of Jena, 1958.
 \bibitem{Gor2}
A. K. {Gorbatsievich}, {Ho Si Mau Tchuc},  and  E. {Schmutzer}, \emph{Acta Phys.
Pol., B} \textbf{ 27} (1996) 1991.
\bibitem{Pitjeva}
 E. V. Pitjeva,  \emph{Astronomy Letters} \textbf{31} (2005b) 940.
 \bibitem{Sanders}
 A. J. Sanders, G. T. Gillies, and E. Schmutzer, \emph{Ann. Phys. (Berlin),} \textbf{522} (2010) 861.
\bibitem{Pound1}
         R.V.~Pound, and G.A.~Rebka,   \emph{Phys.  Rev.  Lett.}  {\bf  4} (1960)  337.
\bibitem{Pound2}
         R.V.~Pound, and J.L.~Snider, \emph{{Phys. Rev.}}, {\bf B140} (1965) 788.
\bibitem{Will1}
Will,\,C.\,M., \emph{Theory and experiment in gravitational physics}
(Cambridge University Press, Cambridge, U.K.; New York, U.S.A.,
1993, 2nd edition).
\bibitem{SchmuExact1}
 E. Schmutzer, \emph{Ann. Physik.} \textbf{4} (1995) 251.
\bibitem{Nato} A. K. Gorbatsievich, in: \emph{The Gravitation Constant: Generalized Gravitational Theories and Experiments,   NATO Science Series} (eds. De Sabbata, T. Gillies, and V. Melnikov, Kluwer  Academic Publishers, Dordrecht, Boston, London, 2004) p. 192.
\bibitem{Schmu16}
E. Schmutzer, in: \emph{Schriften der Sudetendeutschen Akademie der Wissenschaften und K\"unste,}
\textbf{31}  (M\"unchen, 2011) p. 123.
\bibitem{Bekenstein}
 J.D.        Bekenstein,  \emph{Phys. Rev. }{\bf D15}, (1977) 1458.
\end{thebibliography}
\end{document}